\documentstyle[amssymb,aps]{revtex}

\begin{document}
\author{A.Zh. Muradyan and G.A. Muradyan}
\address{Department of Physics, Yerevan State University, 1 Alex Manookian,\\
Yerevan, 375049\\
Armenia\\
e-mail: muradyan@server.physdep.r.am}
\title{Decreasing critical temperature of gas BEC in spatially periodic potential
and relevance to experiments treated by Mott-Hubbard model.}
\maketitle

\begin{abstract}
It is shown that the critical temperature of gas Bose-Einstein condensation
decreases in deepening periodic potential, in contrast to common regularity
in a separate potential well. \ The physical explanation of this phenomenon
is given. \ Characteristic scale of potential energies decaying the critical
temperature is the quantum recoil energy of periodic potential. \ 

The theory represents an alternative and direct approach to the experimental
results (C.Orzel et al Science {\bf 291}, 2386 (2001); M.Greiner et al,
Nature {\bf 415,} 39 (2002)) obtained with BEC in optical lattices and
treated as the phase squeezing or Mott transition processes.
\end{abstract}

\section{Introduction}

\ One of the main characteristics of a BEC is the critical temperature of
condensation \cite{1}. \ As is well known, for free ideal gas (in
thermodynamic limit) it can be simply determined and is presented by formula 
\begin{equation}
T_{co}=\frac{2\pi \hbar ^{2}}{mk_{B}}\left[ \zeta \left( \frac{3}{2}\right) %
\right] ^{-2/3}\left( \frac{N}{V}\right) ^{2/3},  \label{1}
\end{equation}
where $m$ is particle mass, $k_{B}$ is the Boltzmann constant, $\zeta \left(
x\right) $ is the Riemann function, $N$ is the number of gas particles in
volume $V$ . \ This formula has clear interpretation that the condensation,
as a macroscopic quantum phenomenon, starts at the temperature when the de
Broglie wavelength $\lambda _{c}=\sqrt{\left( 2\pi \hbar
^{2}/mk_{B}T_{co}\right) }$ of particle's thermal motion becomes same order
quantity with the distance among the neighboring particles $\left(
V/N\right) ^{1/3};$ it ensures the existence of perceptible overlapping of
neighboring particle wavefunctions(wave-packets), thus making possible the
manifestation of boson-inductive nature of particles.

In real situations, when one however has neither ideal nor free gas, aroused
the natural question how much and in what direction changes the critical
temperature\ compared with the one given by Eq.\label{1}(\ref{1}),
conditioned respectively by interparticle interactions and by existence of
trapping external fields (potentials). \ Interparticle interactions, having
crucial importance for dynamic properties of condensate, such as stability,
excitations and so on, have, fortunately, feeble influence on critical
temperature of gas condensation \cite{1}, \cite{2} (note, that just this
feature allows us to use the ideal gas model for critical temperature
calculations in the following). The influence of the existence of trapping
potential has been studied rather frequently and the result about the
critical temperature can be qualitatively understood by means of Eq.(\ref{1}%
) or reasonings connected with its physics. Actually, the
existence(deepening) of trapping potential directly increases the mean
density of gas relative to the free state and thereby increases the critical
temperature too. \ Here we want to stress, however, that this reason and
consequent growth of critical temperature is relevant to the single-well
potential and can not be directly implemented for multi-well potentials, in
particular for optical lattices\cite{3}, which attract a great deal of
attention \ thanks to fascinating eventual applications such as matter-wave
transport \cite{4} and diffraction \cite{5}, quantum logic \cite{6}, etc. \
The reason, that in multi-well cases we should exercise some caution is that
while in single-well case the deepening of potential leads only to drawing
nearer all the atoms in the trapping potential, in multi-well potentials the
reverse tendency exists too. \ With the deepening of potential, all the
particles that have preliminarily been in barrier-type regions, fill the
neighbor wells, and thereby increasing the atomic concentration in well-type
regions, decrease it in barrier-type regions. \ In optical lattices, in
addition, the mean $\ $atomic density \ is conserved. \ 

The existence of \ two opposite tendencies for the gas density, the
thickening in well-type regions and respective rarefying in intermediate
barrier-type regions(see Fig.\ref{Fig.1}), demands a separate considerations
of the question \cite{7}, even more so it may be expected that the critical
temperature of condensation in periodic in space potential must be a
decreasing one depending on the depth of potential, that is it must be
straightly opposite to the behavior shown in case of single-well potential.
\ To understand the footing of this assertion one should note that the
number of atoms in barrier-type region in general is less than in well-type
region, and besides, the redistribution of atoms between barrier-type and
well-type regions can be represented as substraction of some (definite)
number of atoms from a barrier-type region and their addition into a
neighbor (left-hand side or right-hand side, it doesn't matter) well-type
region. But substraction of a definite quantity from a smaller number is
more essential for itself than the addition of the same quantity to the
larger (or equal) number! \ That is, the decreasing of the wave-function
overlapping in barrier-type regions must be more essential for the total gas
condensation, than the overlapping increasing \ for the well-type regions,
and as a consequence it would lead to mentioned decreasing of the critical
temperature.

In this paper we will afford a quantitative justification of the mentioned
assertion about the decaying behavior of condensation critical temperature
(and as a consequence, of the number of condensed particles) for both
deepening and/or expanding periodic in space potentials. The form of
periodic potential, which will be utilized to this end, will be the
biparabolic form \cite{8}. \ On top of that we shall present a simple and
convincing physical explanation, based on regularities of zonal structure of
energy spectrum, why the depth-dependence of critical temperature in
standing wave trapping potential must be just the contrary to the case of
single-well trapping potential.

We are coming to the problem as in ordinary statistical mechanics, that is,
via the familiar relation between the number of particles (atoms),
temperature and chemical potential\cite{1}. The potential is assumed varying
only in one direction, meaning that the atomic gas is free in other two
directions. \ For the energy of interacting degree of freedom the
repeated-zone scheme is used, where the quasimomentum plays the same role as
the ordinary momentum plays for free degrees of \ freedom. \ A direct
consequence of obtained for critical temperature regularity, the eventual
diminishing of the number of condensed atoms in the deepening potential,
also is elucidated:. In addition we calculate the number of atoms in
separate energy zones as a function of potential depth in order to enlarge
the comparison with the single-potential case. \ Finally we work out the
dependance of critical temperature on space-period of potential and find it
to be decreasing too.

The paper is organized as follows: in Sec.II we shortly represent periodic
in one dimension, so-called biparabolic potential and Bloch-state solutions
for a particle there. \ In Sec.III and IV we outline the theoretical
approach, illustrate the results of numerical calculations about the
critical temperature of condensation, number of condensed particles and
population of energy zones. \ Section V contains some comments and
conclusions.

\section{ Biparabolic potential and dispersion relation for a particle there}

In order to gain an overall picture, let us start and carry the exposition
in terms of interactions potential, not concretizing, in general, the
mechanisms of interaction. \ However recalling that at the present far
off-resonant standing wave laser radiation is used to trap condensed dilute
atomic gases, we will also write down the expressions of potential depth and
period for this special case$.$ \ All the numerical simulations will be
performed, having in view just this potential.

Relevant to problem Schr\H{o}dinger's stationary equation in dimensionless
notations is 
\begin{equation}
\frac{d^{2}\Psi (Z)}{dZ^{2}}+\left[ W-U(Z)\right] \Psi (Z)=0,  \label{2}
\end{equation}
where $Z,W$ and $\ U(Z)$ respectively are dimensionless coordinate along the
periodic potential, total energy relevant to this degree of freedom, and
potential energy of the particle (atom) there.\ They are related with the
physical coordinate $z,$ energy $E$, and potential energy $V(z)$ by means of
the following formulas: 
\begin{equation}
Z=2\pi z/l,\text{ \ }W=E/E_{r},\text{ \ }U(Z)=V(z)/E_{r},  \label{3}
\end{equation}
where $l$ is the space period of potential, $E_{r}$=$\left( 2\pi \hbar
\right) ^{2}/2ml^{2}$ is the quantum recoil energy of periodic potential, $m$
is the particle (atomic) mass.

As a model of periodic field we choose the biparabolic form\cite{8}, 
\begin{equation}
U(Z)=\frac{1-(-1)^{m}}{2}U+(-1)^{m}\Bbbk (Z-m\pi )^{2},  \label{4}
\end{equation}
with $(m-1/2)\pi \leqslant Z\leqslant (m+1/2)\pi $ for any integer $m=0,\pm
1,\pm 2,....$ Here $\ U$ is the depth of potential, $\Bbbk =2U/\pi ^{2}$ is
a coefficient assuring the continuity of potential, consisting of series of
truncated and turned parabolas, as is shown in Fig.\ref{Fig.2} by a solid
line. The potential energy is counted from the bottom of the periodic wells.

Note that this potential can also be regarded as a good approximation to a
potential created for a two-level atom by a standing wave of laser
radiation, far off-resonant with optical transition. \ This relevant
standing wave potential is depicted in Fig.\ref{Fig.2} by a dashed-line, and
then one has 
\begin{equation}
2\pi /l=2k,\text{ \ \ }V(Z)=4\left| dE_{L}\right| ^{2}/\hbar (\omega -\omega
_{0}),  \label{5}
\end{equation}
where $k=\omega /c$ is the laser wavenumber, $d$ is the dipole matrix
element of optical transition of frequency $\omega _{0},$ $E_{L}$ is the
amplitude of laser electric field for both traveling waves, composing the
standing wave.

Bloch-solutions of Eq.(\ref{2}) with potential (\ref{4}) results in a
following dispersion relation: 
\begin{equation}
\cos \left[ 2\pi P\right] =1+2G_{11}(W)G_{22}(W)=1-2G_{12}(W)G_{21}(W),
\label{6}
\end{equation}
where $P$ is the particle's quasimomentum in units $2\pi \hbar /l$( =$2\hbar
k$ in laser standing wave case), and the normalized energy $W$ is presented
in the second equation of (\ref{3}). Present in (\ref{6}) functions $%
G_{ij}(W)$ $(i,j=1,2)$ are expressed via the linearly independent solutions $%
u_{1}(Z)$ and $u_{2}(Z)$ in the well-type region, $\widetilde{u}_{1}(Z)$ and 
$\widetilde{u}_{2}(Z)$ in the neighbor barrier-type region, and their first
derivatives at the connective point $Z=\pi /2$ as 
\begin{equation}
G_{ij}(W)=\left[ u_{i}(Z)\frac{d\widetilde{u}_{j}(Z)}{dZ}-\widetilde{u}%
_{j}(Z)\frac{d\widetilde{u}_{i}(Z)}{dZ}\right] _{Z=\pi /2},\text{ }i,j=1,2%
\text{ },  \label{7}
\end{equation}
\begin{equation}
u_{1}(Z)=\exp \left[ -\frac{\sqrt{\Bbbk }Z^{2}}{2}\right] \Phi (\alpha ,1/2;%
\sqrt{\Bbbk }Z^{2}),  \label{8}
\end{equation}
\begin{equation}
u_{2}(Z)=Z\exp \left[ -\frac{\sqrt{\Bbbk }Z^{2}}{2}\right] \Phi (\alpha
+1/2,3/2;\sqrt{\Bbbk }Z^{2}),
\end{equation}
\begin{equation}
\alpha =\frac{1}{4}\left[ 1-\frac{W}{\sqrt{\Bbbk }}\right]  \label{10}
\end{equation}
and $\widetilde{u}_{1,2}(Z)$ are obtained from $u_{1,2}(Z)$ respectively by
means of substitutions $Z^{2}\longrightarrow -i(Z-\pi )^{2},$ $\ \alpha
\longrightarrow \beta =\frac{1}{4}\left[ 1-i\frac{W-U}{\sqrt{\Bbbk }}\right]
.$ \ Here $\Phi (...,..;...)$ is the confluent hypergeometric function. \
Let us note that the functions $\widetilde{u}_{1,2}(Z)$ also are real
functions of coordinate $Z$ and physical parameters, although they contain
imaginary argument $-i(Z-\pi )^{2}$ and complex parameter $\beta $\cite{8}.

\section{Critical temperature and number of condensed atoms as a function of
depth of periodic in space potential}

Our analysis of ideal gas critical temperature and number of condensed
particles (atoms) will be worked out by the familiar relation between the
mean number of particles, temperature and chemical potential\cite{1},
turning from the summation over energy spectrum to integration over the
momenta. For the energies corresponding to\ along the potential's
periodicity\ degree of freedom, it is convenient use the repeated-zone
scheme of energies where each value of energy specifies a single value of
quasimomentum (see Fig.\ref{Fig.3}). Quasimomentum naturally plays the role
of ordinary momentum. In this representation the mentioned relation can be
written in form

\begin{equation}
2\frac{V}{l^{3}}\sum\limits_{i=1}^{\infty }\int\limits_{(i-1)/2}^{i/2}dP\int
\int\limits_{-\infty }^{\infty }\frac{d\overrightarrow{P}_{\perp }}{\exp %
\left[ \frac{W_{\perp }+W-\mu }{k_{B}T}\right] -1}=N.  \label{11}
\end{equation}
$V$ and $N$ are the gas volume and the mean number of particles. The
momentum $d\overrightarrow{P}_{\perp }$ (in the free-motion $X0Y$ plane) is
normalized, as the quasimomentum $P,$ in $2\pi \hbar /l$ units ($=2\hbar k$
in laser standing wave case), and the energetic quantities $W_{\perp },$ $%
\mu $ and $k_{B}T$ are normalized in units $E_{r}$ ($=(2\hbar k)^{2}/2m$ in
laser standing wave case) as the energy $W.$ \ Note that in normalized form $%
W_{\perp }=\overrightarrow{P}_{\perp }^{2}.$ \ Integration over the $P$%
-momenta is divided into integration over consequent zones $i=1,2,...$ . \
And finally, integration over the negative values is changed into
positive-values too, due to which the coefficient $2$ has appeared in
left-hand side of (\ref{11}). \ It is to be noted here that the state
density is zero at $P=0$ (as for a free gas) and the twofold repeating of
the point $\ P=0$ in (\ref{11}) does not matter.

Performing the integration over $\overrightarrow{P}_{\perp }$ (free degrees
of freedom), instead of (\ref{11}) we arrive at 
\begin{equation}
-2\frac{(\pi /d)^{3}Vk_{B}T}{\pi ^{2}}\sum\limits_{i=1}^{\infty
}\int\limits_{(i-1)/2}^{i/2}dP\ln [1-\exp [\frac{\mu -W(P)}{k_{B}T}]]=N.
\label{12}
\end{equation}

Because of the zero state-density at $P=0,$ $\overrightarrow{P}_{\perp }=0$
(ground state), the critical temperature $T_{c}$ and the number of condensed
particles $N_{c}$ are afforded to be determined in full analogy with the
free ideal gas, that is to consider (\ref{12}) as a determination of $\mu $
for a given $N,$ if $T\geqslant $ $T_{c},$ but for lower temperatures $%
T\leqslant $ $T_{c}$ to fix the value of $\mu (T\leqslant T_{c})=\mu
(T=T_{c})$, and consider (\ref{12}) as a relation for determination of the
number of noncondensed particles $N_{nc}.$ \ For the last case, when the gas
is cooled lower than the temperature of condensation, the number of
condensed atoms is \cite{9} 
\begin{equation}
N_{c}=N-N_{nc}.  \label{13}
\end{equation}

Equations (\ref{12}), (\ref{6})and (\ref{13}), the second of which
determines a one-valued implicit function $W(P)$ in repeated-zone picture,
are the basis of the reminder of this paper.

Let us now present the scheme of calculations, which will be used here. \ To
determine $T_{c}$ we first insert the values of interest of $N,$ $V$ and $l$
into (\ref{12}) and consider it as a relation, which determines the chemical
potential $\mu $ as a function of temperature $T$ for a given dispersion
relation $W(P)$. \ Since the seeking functional dependence $\mu (T)$ can be
obtained only by means of numerical integration in(\ref{12}) and $\mu $ sits
in integrand, we preliminarily cast the $T(\mu )$ dependence and hereupon
turn to $\mu (T)$ function numerically, using the one-valued behavior
dependences. In its turn to obtain the $T(\mu )$ dependence we give values
to $\mu $ and performing the numerical integration in (\ref{12}) find the
corresponding values of $\ T$. \ The stage of integration demands of course
to fix the depth of the potential too, to have the $W(P)-$dependence. \ The
latter we get by reversing the dispersion relation (\ref{6}) numerically,
where the repeated-zone picture is very convenient.

Construction of the $T(\mu )$ dependence in practice starts from some value
of $\mu ,$greater than the numerical value of the particle energy $W_{\min }$
for the chosen potential, and gradually move to smaller values of $\mu .$\
The possible minimal (having solution of (\ref{12}) ) value of $\mu $ is $%
\mu =W_{\min },$ and corresponding to it value of temperature, as solution
of (\ref{12}), just specifies the critical temperature $T_{c}.$ \ Some
graphs of the $\mu (T)$ function for five values of potential depth $U$,
constructed in mentioned above way, are depicted in Fig.\ref{Fig.4}. \ In
accordance with the described procedure of calculations, the value of
critical temperature for each graph-conditions is determined by the
corresponding left-hand side border point of graph. \ Now let us concentrate
attention on the fact that the graphs from right to left correspond to
lifting values of potential depth, which immediately shows the anticipated
in Introduction behavior: \ the critical temperature of ideal gas
condensation decreases as the potential depth increases.

It should be noted specially, that the determination of the border points of
graphs (and corresponding values of $T_{c}$) demands a great caution of
numerical integration procedure, since the integrand at $\mu \preccurlyeq
W_{\min }$ (near-border values) rapidly increases near the lower edge of the
first energy zone, that is near $P=0,$ where $W(P=0)=W_{\min },$ $\mu
-W(P=0)\rightarrow 0,$ $1-\exp \left[ (\mu -W(P=0))/k_{B}T\right] \ll 1$,
and respectively $\ -\ln \left[ 1-\exp \left[ (\mu -W(P=0))/k_{B}T\right] %
\right] \gg 1.$ \ Fortunately, the theoretical value of the border (maximal)
value of $\mu $ is determined by exact condition $\mu _{\max }=W_{\min },$%
which can be fulfilled as precisely as it is possible within the range of
computer capability. The minimal accuracy for the (dimensionless) difference
has been chosen $10^{-8}.$ \ Besides, the procedure of integration in (\ref
{12}) was replaced by corresponding summation, dividing the first zone
quasimomentum range into $10^{3}$ equal parts, and the first of them, which
is about the $P=0,$ already into $10^{4}$ equal parts. As a result, the time
consumption of our PC for each point on the presented graphs was about 30-40
minutes.

In Fig.\ref{Fig.5} we explicitly present the dependance of gas critical
temperature on the depth of periodic potential. \ It has a monotonically
decreasing nature everywhere, rather minor in small-depth range and rapid in
intermediate-depth range of the potential. \ In the range of greater depths
the rate of decreasing , naturally, slows down. \ 

The number of condensed atoms , as it directly follows from the mentioned
behavior of critical temperature, decreases with the growth of potential
depth too. \ In addition, since the critical temperature monotonically tends
to zero as a function of $U$, for arbitrary chosen low temperature $T$ there
is always some potential depth, for which the critical temperature becomes
equal to $T,$ and starting from which no more condensation takes place. \
This behavior of the number of condensed atoms is shown in Fig.\ref{Fig.6}.
\ The point where the graph touches the depth axis, just corresponds to the
peculiar value of potential depth, for which the critical temperature
becomes equal to the chosen temperature of the gas ($0.1T_{0c}$ in this
case). \ For dipper potentials ($U>13$) the critical temperature is lower
then the gas temperature and because of it no condensation takes place.

Before proceeding to other results obtained, let us turn to other language
of presentation, explaining why the number of condensed particles decreases
with the increasing depths of the periodic potential, which is the
equivalent of critical temperature decreasing. \ This representation is
based on the behavior of energy spectrum for increasing depths of potential
and gives, perhaps, more detailed and transparent explanation, than the
previous one, based on behavior of particle density. \ It easily explains
the studying regularity also in single-potential case. \ Indeed, in a
single-potential the particle has discrete set of energy levels $E_{n}$ $%
(n=0,1,...)$ of bounded states (only they are relevant to the problem under
hand). The probability of population for each excited level, relative to
population of the ground energy level, is determined by the Boltzmann factor 
$\exp \left[ -E_{n}/k_{B}T\right] $ in thermally equilibrium state. \ The
energy $E_{n}$ of each excited energy level can be also interpreted as the
distance from ground energy level, choosing without loss of generality the
energy of the latter zero. \ The deepening of a single potential, as is well
known increases distances between energy levels and therefore, between each
of them and ground energy level. \ As a result, the population of all
excited energy levels decreases. \ In addition taking into account that the
total probability of level occupation (population) remains the same unity,
we immediately arrive to the result, that the occupation probability of the
ground energy level must increase due to this deepening of the potential. \
But the condensed gas is just the particles in ground energy level and
therefore, the deepening of a single-potential results in increasing of
condensed particles, or which is equivalent, in increasing of condensation
critical temperature.

In periodic-potential case the situation is, nevertheless, different. \ The
spectrum of a particle has a zonal structure. And while the deepening of
potential removes from each other the energy bands, it simultaneously
narrows all these bands, in particular the first one, which plays the
dominant role in phenomenon of condensation. \ In order to easily understand
the resultant seeking behavior in this situation, it is convenient to
present the first energy band as a collection of discrete energy levels ( in
case of finite, even arbitrary large, number of repeated potential wells
this approach is absolutely correct ). \ To the phenomenon of condensation
is relevant, by definition, only the ground state. \ Repeating the above
discussion about the behavior of level occupation probabilities for
deepening potentials we see, that the occupation probabilities increase in
all excited levels in the first energy band and decrease in all upper energy
bands. \ But since the condensation appears at low temperatures, when the
low-energy levels are mainly populated, the increase of population in
excited first-bound levels plays more important role, than the decrease of
population in upper energy levels. Therefore the increase of potential
depth, at least for low temperatures, must result in increment of population
in the total excited states, including all the excited levels of the first
energy band and all excited energy bands. \ This implies that one has
decrement of population in ground energy level, relevant to the state of
condensation. \ So, regularity of the ground state population for a series
of deepening periodic potentials is opposite to the case of a single
potential: the number of condensed particles is decreasing and the critical
temperature of condensation is respectively decreasing. \ It is not needless
to note, that the presented discussion, based on the behavior of energy
spectrum and relevant Boltzmann function can help us understand
qualitatively the behavior of number of condensed atoms, or equivalently,
the critical temperature in other potential forms too, particularly in
double well potential.

In order to give somewhat auxiliary elucidation to the problem under hand,
we have calculated the populations of each energy band as a function of
depth of periodic potential for some, over-critical, temperature, when not
only the first energy band is effectively populated. \ The results are
presented in Fig.\ref{Fig.7}(point lines). \ As is seen, the population of
the first energy band increases monotonically as the depth of periodic
potential increases, while the populations of the all excited energy bands,
vice versa, decrease. \ It \ also gives an opportunity to see, that these
regularities for energy band populations in periodic-wells potential
precisely coincide with the regularities of the level-populations for
single-well potential. \ This implies, in particular, that the growing
behavior of the critical temperature of condensation would have been
obtained in periodic potential case, if one had taken as a ground level the
first energy zone totally, which is however incorrect by definition, even
for very deep potentials.

In envisioned circumstances some interest may also present the total number
of particles, trapped by the periodic potential. That are the particles,
energies of which are less than the depth of potential. \ Because of a zonal
structure of energy spectrum, in general we should expect a stepwise
dependence of this number on the potential depth. \ The results are
illustrated in Fig.\ref{Fig.8}. \ At the origin (point $a_{1}$) with $U=0$,
of course there are no trapped particles. The piece $a_{1}b_{1}$ corresponds
to situation when the rising potential captures more and more of the first,
narrowing at that, energy band, naturally increasing the number of trapped
particles. \ Since the energy bands are distanted from each other by energy
gaps, the pieces corresponding to the second, third, etc. bands are $%
a_{2}b_{2}$, $a_{3}b_{3},etc$ (the upper range is not presented), distanted
from each other by intermediate pieces $b_{1}a_{2},$ $b_{2}a_{3},$ etc. All
the regularities of the graph, such as the rate of increasing, saturation,
etc., are in agreement with the presented picture, viewing the distribution
of quantum states within the energy bands and their thermal population as a
function of potential depth.

\section{Critical temperature as a function of period of periodic potential}

Another side of interest presents \ the dependence of critical temperature
of condensation on the period of periodic in space potential. \ In case of
laser-induced \ standing wave potentials the variation of the period can be
obtained varying the angle between two waves, composing the standing wave.
Denoting by $\theta $ the half of supplementary angle \ between these two
wave vectors (see Fig.\ref{Fig.9}), we get $k_{SW}=2k\cos \theta $ and
respectively $l=\pi /k\cos \theta .$ $\theta =0$ corresponds to
counterpropagating case, where the period is minimal. \ We have found
convenient to present the $l$-dependence of critical temperature as a $%
\theta $-dependence. \ The scheme of calculations is the same with the one
described in previous section. \ Numerical calculations had been worked out
for two ($(a)U=1$\ and \ $(b)U=6$ ) depths of potential and the results are
plotted in Fig.\ref{Fig.10}. \ The critical temperature, as is seen \
decreases as the period of potential increases, monotonically approaching
zero.

We have not been able to understand this result on the base of picture
about\ the overlapping of particles' material waves (see Introduction), but
it is easy to do, using the picture of energy spectrum and thermal
population of energy levels there. \ Really, the spatial expansion of
periodic potential results in narrowing of all energy bands and
simultaneously in widening of the energy gaps. \ But it is the same
regularity as it was for the deepening potential, and therefore the
resultant functional dependance also is the same: the critical temperature
decreases as the spatial period increases.

Of course, an analogous regularity exists also for the number of condensed
atoms, but we do not see the necessity of presenting this concrete form.

\section{Concluding remarks}

So, the critical temperature of condensation and respectively the number of
condensed particles in a trapping periodic potential must be less than it is
in free state. \ 

An important question is, of course, how much ''dangerous'' are these
results for experimental situations, typical for condensates trapped in
optical standing wave potentials. \ The critical value of thermal energy of
free gas condensation in typical range of densities $10^{11}-10^{12}$ $%
cm^{-3}$, as is well known is far subrecoil ( is on the order of one percent
or less). \ This implies that such a gas may be trapped, in principle, by
means of an optical standing wave potential with the depth of the same order
of magnitude. \ Looking at the Figs.\ref{Fig.5} and \ref{Fig.6} we see that
this subrecoil range of potentials lies close to the origin of horizontal
axis, where the lowering of graphs ( critical temperature \ or/and number of
condensed particles) is unessential. Moreover, the characteristic parameters
of condensation in practice stay unchanged up to depths $U/E_{r}\approx 1$,
that is in so called shallow potentials (note that our scale of recoil
energy $E_{r}=(2\hbar k)^{2}/2m$ is four times greater than the usually
using scale $\hbar ^{2}k^{2}/2m$). \ However, in deeper periodic potentials
with $U/E_{r}\approx 5\div 12$(in usual scale $\approx 20\div 50$), which
are exploited for strong localization of super-cooled condensed gases within
separate wells, the critical temperature of condensation falls about one \
order and the gas can get out from the condensation state.

The comparison with the real experimental situation involves another
important parameter too: it is the time, especially in cases when the gas is
filled into the standing wave potential already being in condensed state
(obtained in magneto-optical traps). \ The matter is that the above
presented theory pertains to thermally equilibrium gas in the given external
potential field. \ The gas, finding itself in the periodic field,
preliminarily is not in that state, but it would tend to that state in the
course of time. \ The dynamics of approaching to the thermally equilibrium
state is, in general, complicated and requires a separate and accurate
consideration, but on qualitative level it may be expected, that two
processes would determine the transfer of gas into statistical-mechanical
equilibrium state in the case under consideration. \ The first one is the
process of interparticle collisions, leading to thermal equilibrium, and the
second one is the tunneling of particle through the potential barrier
between neighboring potential wells, providing the establishment of zonal
structure for particle energy spectrum. \ Denoting their characteristic
times by $\tau _{col}$ and $\tau _{tun}$ respectively, it will be expected
the ultimate thermally equilibrium distribution (with lower critical
temperature and less number of condensed particles) should be attained in
times $t\geqslant \tau _{col},$ $\tau _{tun}.$

Let's note that from the viewpoint of our results the distortion of gas
BEC-state in a deepening periodic potential and its consequent revival in a
shallowing one, observed experimentally in \cite{10, 11} as a loss and
revival of phase, may be accounted simply as a respective transfer to
noncondensed state and consequent return into the condensed one, as is
predicted by the above presented theory for a rather slowly deepening and
shallowing periodic potential. All the more, that the totally distorting
potential depths$\ U/E_{r}^{o}=22$ ($=5.5$ in our $E_{r}$-scale) in \cite{10}
and $44$ ($=11$ in our $E_{r}$-scale) in \cite{11}, lie in the region of our
curve of Fig\ref{Fig.5}, where the condensed state must be depressed\ in
practice. Plus, that in three-dimensional standing wave case \cite{10} the
analogous to Fig\ref{Fig.5} curve must decay more rapidly than our,
calculated for one-dimensional standing wave.

\ The detailed elusidation of results of these experiments in view of
presented here theory will be accomplished and presented elswhere.

\section{Acknowledgments}

We are grateful to T.H.Bergeman for useful remarks. \ This paper is
supported by the Armenian State Scientific Grant No. 0888.

\begin{center}
\begin{figure}[tbp]
\caption{Scheme of gas spacial distribution for $(a)$ a free state, $(b)$
intermediate depth and $(c)$ great depth of periodic potential. \ The mean
concentration of gas stays unchanged with the potential depth evolution.}
\label{Fig.1}
\end{figure}
\begin{figure}[tbp]
\caption{Biparabolic form of potential (solid line), used in our
calculations. \ For comparison is also presented the respective sinusoid
(dashed line) of laser standing wave potential. }
\label{Fig.2}
\end{figure}
\begin{figure}[tbp]
\caption{Repeated -zone scheme of single particle energy spectrum, which has
been used in our calculations. \ The picture is for the case $U=1$ (in
recoil energy units).}
\label{Fig.3}
\end{figure}
\begin{figure}[tbp]
\caption{The temperature dependance of chemical potential $\protect\mu %
(N/V,T)$ for different depths of potential $\ ((a)U=0,$ $(b)0.5,$ $(c)2.5,$ $%
(d)7,$and $(e)12)$ and for $N/V=10^{11}cm^{-3}.$ \ Note that the values of $%
\protect\mu $(vertical axis) are presented relative to minimal energy $%
E_{\min }$ of particle in periodic potential. \ Both axis are normalized by
recoil energy $E_{r}$.}
\label{Fig.4}
\end{figure}

\begin{figure}[tbp]
\caption{The behavior of critical temperature $T_{c}$ of ideal gas
condensation as a function of depth $U$ in external periodic potential. \
The vertical axis $T_{c}$ is normalized by critical temperature \ $T_{c0}$
of free gas (with $N/V=10^{11}cm^{-3}$), horizontal axis $U$ by recoil
energy $E_{r}$, as it is done along all the paper. \ The discovery of this
behavior is the main result of the presented study.}
\label{Fig.5}
\end{figure}

\begin{figure}[tbp]
\caption{The periodic potential depth dependence of the number $N_{c}$ of
condensed particles(atoms). \ The gas temperature is $T=0.1T_{c0},$ that is
ten times below the free gas critical temperature $T_{c0}$ of condensation
(in case under consideration $k_{B}T_{c0}=0.002163774E_{r},$ or equivalently 
$T_{c0}=1.03819\ast 10^{-8}K$ if one is concerned with $D2$ optical
transition in sodium atom with $\protect\lambda =5896\stackrel{\circ }{A}$).
\ The vertical axis is normalized by the total number $N$ of particles in
the gas.}
\label{Fig.6}
\end{figure}
\begin{figure}[tbp]
\caption{The periodic potential depth dependence of the numbers $N_{zone}$
of particles, populating first, second, and third zones respectively. \ The
latter are given in units of total number $N.$ \ The gas temperature is
chosen $k_{B}T=1.5$ and is far over the critical value of condensation. \
The points represent the numbers, irrelevant to the fact are the particles
trapped by the periodic potential or not, whereas the dashed line represents
the number of trapped particles (that is particles with $E<U$). \ In deep
potential regions, when the zone is entirely within the potential, both
quantities coincide. \ As is seen (points) the population of the first zone
monotonically increases as potential becomes deeper, while the population of
the others decreases, in full analogy with the regularity of energy- level
populations in single -well potential case. }
\label{Fig.7}
\end{figure}
\begin{figure}[tbp]
\caption{The periodic potential depth dependence of the number $N_{tr}$ of
particles, trapped by the potential. \ The deepening of the potential
includes more and more number of zones, thereby increasing the number $%
N_{tr}.$ \ The stepwise behavior of the dependence is conditioned by the
zonal structure of energy spectrum (for details see the text). }
\label{Fig.8}
\end{figure}
\begin{figure}[tbp]
\caption{The geometry of two laser waves, creating the standing-wave
periodic potential. $\protect\theta =0$ corresponds to the
counterpropagating case. \ Variation of $\protect\theta $ changes the
spatial period, but keeps the depth constant.}
\label{Fig.9}
\end{figure}

\begin{figure}[tbp]
\caption{The behavior of critical temperature \ $T_{c}$ of ideal gas
condensation as a function of angle $\protect\theta $ (see previous Figure).
\ The spatial expansion of periodic potential, as the deepening decreases
the value of critical temperature of condensation.}
\label{Fig.10}
\end{figure}
\end{center}


\begin{references}
\bibitem{1}  K.Huang, Statistical Mechanics (Wiley, New York, 1963), Ch.12
\S 3, Ch.13 \S 5.

\bibitem{2}  V.Bagnato, D.E.Pritchard and D.Kleppner, Phys. Rev. A {\bf 35,}
4354 (1987).

\bibitem{3}  K.Berg-Sorensen and K.Molmer, Phys. Rev. A, {\bf 58,} 1480
(1998); J.Javanainen, Phys. Rev. A {\bf 60,} 4902 (1999); J.C.Bronski,
L.D.Carr, b.Deconinck, and J.N.Kutz, Phys. Rev. Lett. {\bf 86}, 1402 (2001);
A.Trombettoni and A.Smerzi, J. Phys.B: At. Mol. Opt. Phys. {\bf 34}, 4711
(2001); {\it ibid}, Phys Rev.Lett. {\bf 86}, 2353 (2001); D.Jaksch,
V.Venturi, I.J.Cirac, C.J.Williams, and P.Zoller, cond-mat/0204137(2002);
J.H.Denschlang, J.E.Simsarian, H.Haffner, C.McKenzie, A.Browaeys, D.Cho,
K.Helmerson, S.L.Rolston, and W.D.Phillips, J. Phys.B: At. Mol. Opt. Phys. 
{\bf 35}, 3095 (2002).

\bibitem{4}  M.Raizen, C.Salomon and Q. Niu, Physics Today {\bf 50, }30
(1997); C.F.Bharucha, K.W.Madison, P.R.Morrow, S.R.Wilkinson, B.Sundaram,
and M.G.Raizen, Phys. Rev. A {\bf 55}, R857 (1997); B.P.Anderson and
M.A.Kasevich, Science {\bf 282}, 1686 (1998); K.Berg-Sorensen, Y.Castin,
K.Molmer, and J.Dalibard, Europhys. Lett. {\bf 22}, 663 (1993); S.Bernet,
R.Abfalterer, C.Keller, M.K.Oberthaler, J.Schmiedmayer, and and A.Zeilinger,
Phys. Rev. A {\bf 62}, 023606 (2000).

\bibitem{5}  A Zh.Muradyan, Sov. Phys. Izvestia AN Arm SSR, Fizika {\bf 10,}
361 (1975); R.J.Cook and A.F.Bernhart Phys. Rev. A {\bf 18,} 2533 (1978);
A.P.Kazantsev and G.I.Surdutovich, Sov. Phys. Pis'ma JETP {\bf 31,} 542
(1980); A.F.Bernhardt and C.Cohen-Tannoudji; J. Opt. Soc. Am. B {\bf 2,}
1707 (1985); P.Meystre, E.Schumacher and S.Stenholm, Opt. Common. {\bf 73,}
443 (1989); P.J.Martin, P.L.Gould, B.G.Oldaker, A.H.Miklich, and
D.E.Pritchard, Phys. Rev. A {\bf 36,} 2495 (1987).

\bibitem{6}  D.Jaksch, C.Bruder, I.J.Cirac, C.W.Gardiner, and P.Zoller,
Phys. Rev. Lett. {\bf 81}, 3108 (1998); G.K.Brennen, C.M.Cavis, P.S.Jessen,
and I.H.Deutsch, Phys. Rev. Lett. {\bf 82}, 1060 (1999); A.Sorensen and
K.Molmer, Phys. Rev. Lett. {\bf 83}, 2274 (1999).

\bibitem{7}  A.Zh.Muradyan and H.L.Haroutyunyan, Izvestia NAN Armenii,
Fizika {\bf 35,} 3 (2000).

\bibitem{8}  A.Zh.Muradyan, Fizika Tverdogo Tela {\bf (}Rus.{\bf ) 41, }1317
(1999).

\bibitem{9}  L.D.Landau and E.M.Lifshits, Statistical Physics, Part 1
(Nauka, Moscow, 1976), Ch.5 \S 62; R.Balescu, Equlibrium and Nonequilibrium
Statistical Mechanics, Part 1 (John Wiley and Sons Inc., New York-Toronto,
1975), Ch.5 \S 7.

\bibitem{10}  M.Greiner, O.Mandel, T.Esslinger, T.Hansch, and I.Bloch,
Nature {\bf 415,} 39 (2002).

\bibitem{11}  C.Orzel, A.K. Tuchman, M.L.Fenselau, M.Yasuda, and M.A.
Kasevich, Science {\bf 291}, 2386 (2001).
\end{references}
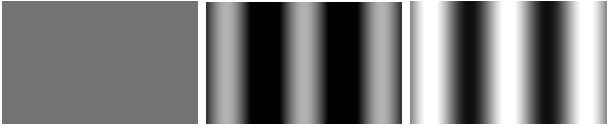
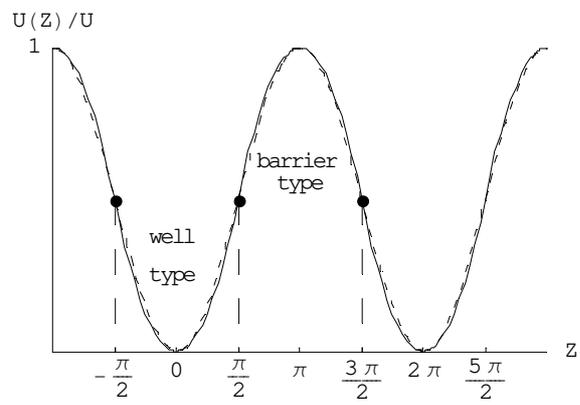
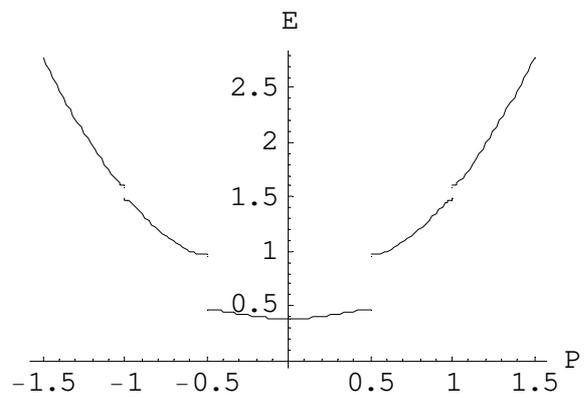
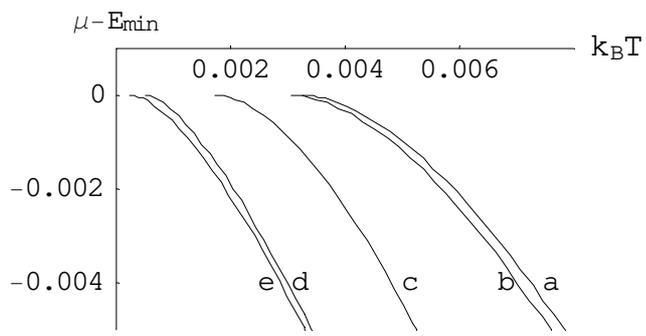
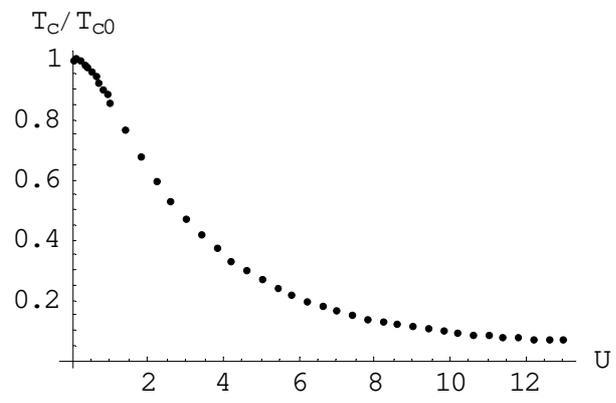
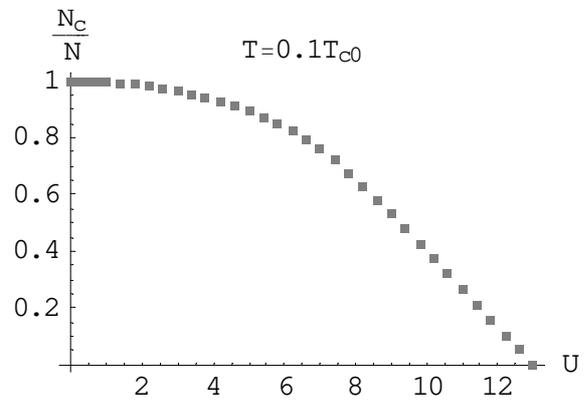
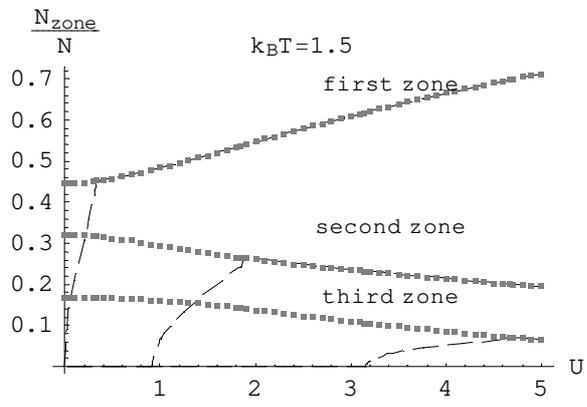
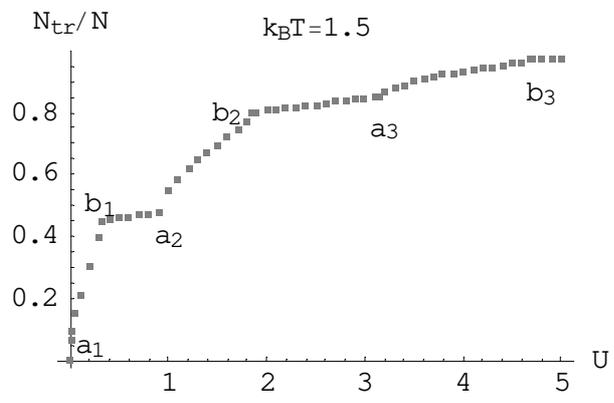
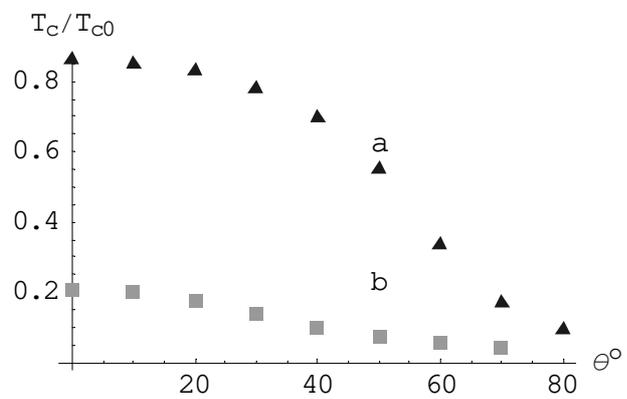
\end{document}